# Scattering Cancellation based Cloaking for the Maxwell-Cattaneo Heat Waves

M. Farhat[1,*], S. Guenneau[2], P.-Y. Chen[3], A. Alù[4] , and K. N. Salama[1]

[1]*Division of Computer, Electrical, and Mathematical Sciences and Engineering, King Abdullah University of Science and Technology (KAUST), Thuwal 23955-6900, Saudi Arabia.*

[2]*Aix Marseille Univ, CNRS, Centrale Marseille, Institut Fresnel, Marseille, France.*

[3]*Department of Electrical and Computer Engineering, University of Illinois at Chicago, Chicago, Illinois 60607, USA.*

[4]*Photonics Initiative, Advanced Science Research Center, City University of New York, New York, NY 10031, USA.*

*In this work we theoretically propose scattering cancellation-based cloaks for heat waves that obey the Maxwell-Cattaneo equation. The proposed cloaks possess carefully tailored diffusivity to cancel the dipole scattering from the object that they surround, and thus can render a small object invisible in the near and far fields, as demonstrated by full-wave finite-element simulations. Mantle heat cloaking is further analyzed and proposed to simplify the design and bring this cloaking*

---

[*] Email: mohamed.farhat@kaust.edu.sa

*technology one step closer to its practical implementation, with promising applications in nanoelectronics and defense related applications.*

## I. Introduction

Ever since the scattering cancellation technique was proposed in 2005 (1), there have been a keen interest of the photonics community and significant studies investigating its intriguing mechanisms and promising applications for different kinds of waves including, but not limited to, microwaves (2), optics (3), elastodynamics (4, 5), acoustics (6-8), thermal (9), or matter (10). Soon afterwards, Pendry and coauthors proposed in 2006 first theoretically (11) and then experimentally, to cover a copper disc of 25 mm radius with a cylindrical coating made of split ring resonators, carefully tailored, to make the medium heterogeneous in a way to curve the path of light in it. In fact, one says that an object is *invisible*, if it does not modify (or little) the wave-field, in which it is embedded (12). One speaks of transparency, if the object itself possesses these properties, and of cloaking, if it is surrounded by a device that offers the same functionality. In this way, invisibility cloaks guide waves around the considered area by creating a hole in the space metric. A related, quite different technique was proposed in the same time by Leonhardt on conformal mapping based cloaking (13). Several other cloaking mechanisms have been put forward in recent years. These range from active cloaking based on anomalous localized resonances (14, 15), homogenization-based cloaking (16, 17), or waveguide theory (18).

When it comes to heat waves, the design of convenient invisibility devices is of paramount importance due to potential applications related to heat exchange, power integration, microelectronics, and even green-building construction or defense (19). Both transformation optics (TO), anomalous resonance (AR), and scattering cancellation techniques (SCT) were applied towards this goal (20-28). In particular, SCT offers simpler designs than the aforementioned approaches, when the size of the objects is smaller than or comparable to the wavelength and the dipole approximation applies (29). In this scenario, isotropic layers of specific diffusivity are sufficient to cancel the first-order scattering and thus significantly reduce the scattering cross-section (or radar cross-section) of the core-shell system. It was even shown that mantle cloaks, i.e. ultra-thin layers, can mimic invisibility for such kind of waves (30-33).

In this work, we propose to generalize the concepts of SCT and mantle cloaking to the field of heat waves obeying the Maxwell-Cattaneo equation (34, 35). The Fourier transfer law of heat gives rise to a parabolic equation that results in instantaneous propagation speeds, which is an unphysical, as is the case with classical mechanics (36). This inconsistency is solved in the latter case by the theory of special relativity (37). Maxwell-Cattaneo law in this sense is a generalization of the Fourier law that results in hyperbolic equations, and therefore finite speeds of propagation. In the following sections, we discuss the details of the cloak design, the physical mechanisms of this scattering cancellation mechanism, and its mantle cloaking realization in the framework of Maxwell-Cattaneo heat waves.

## II. Dispersion Mechanism of Maxwell-Cattaneo Waves

**II.1. Maxwell-Cattaneo Equation**

Thanks to all the aforementioned advantages, we propose, in this work to apply the SCT to heat waves, using the Maxwell-Cattaneo formalism. As a matter of fact, when the Fourier law is used to describe the heat transfer, i.e.,

$$\Phi = -\kappa_0 \nabla T, \qquad (1)$$

with $\Phi$ the local heat flux density (in $Wm^{-2}$), $\kappa_0$ the conductivity of the medium, that can be anisotropic (in $Wm^{-1}K^{-1}$) and $T$ the temperature field (in $K$), it is well-known that this law leads to a parabolic equation for $T$ (the Fourier heat equation). Thus, any initial disturbance in the medium (or object) is propagated *instantly*, due to the parabolic nature of the equation (36). In order to circumvent this apparent unphysical situation, one can use instead the Maxwell-Cattaneo law, which is one of various widely accepted modifications of the Fourier law (35). It takes the form

$$\left(1 + \tau_0 \frac{\partial}{\partial t}\right)\Phi = -\kappa_0 \nabla T, \qquad (2)$$

with $\tau_0$ being the thermal relaxation time (in the order of picoseconds for metals). The additional term $\tau_0 \partial_t \Phi$ accounts for thermal inertia, which avoids the so-called phenomenon of infinite propagation. It is also necessary to add another corrective term that accounts for flux diffusion. The Maxwell-Cattaneo equation, as above-described, takes therefore the final form of a system of coupled partial differential equations (PDEs)

$$\frac{\partial T}{\partial t} = -\nabla \cdot \Phi,$$

$$\tau_0 \left( \frac{\partial \Phi}{\partial t} - \sigma \Delta \Phi \right) = -\Phi - \kappa_0 \nabla T, \tag{3}$$

can be re-written as

$$\tau_0 \frac{\partial^2 T}{\partial t^2} + \frac{\partial T}{\partial t} - \kappa_0 \Delta T - \tau_0 \sigma_0 \Delta \left( \frac{\partial T}{\partial t} \right) = \delta \tag{4}$$

with $\delta$ being the Dirac distribution representing the source term and $\sigma$ accounting for the diffusive phenomena.

### II.2. Dispersion Relation of the Maxwell-Cattaneo Equation

Assuming a time harmonic incident wave, i.e. proportional to $e^{-i\omega t}$, with $\omega$ the angular frequency, Eq. (4) is further simplified to (by omitting the $\delta$ term)

$$\Delta T + \frac{\omega(\omega \tau_0 + i)}{\kappa - i\omega \tau_0 \sigma_0} T. \tag{5}$$

By inspection of Eq. (5), one sees immediately that an effective wavenumber $k_0$ can be defined as $k_0 = \omega \sqrt{(\omega \tau_0 + i)/(\kappa \omega - i\omega^2 \tau_0 \sigma_0)}$. This means that $k_0$ is a complex number for all frequencies, which is markedly different from classical heat waves (Fourier transfer) or even diffusive waves (or diffuse photon density waves, DPDWs) that possess complex wavenumbers in specific frequency ranges. This phenomenon is quite understandable, since one expects new physics, to be at play. Figure 1(b) plots the dispersion relation, i.e., $k_0$ versus the normalized frequency $\omega \tau_0$, for two specific scenarios. In the upper panel of Fig. 1(b), one assumes an extra term (corresponding to DPDWs) in $k_0$, i.e. $k_0 = \omega \sqrt{(\omega \tau_0 + i + \omega \tau_d)/(\kappa \omega - i\omega^2 \tau_0 \sigma_0)}$, with $\tau_d$ the lifetime of photons. The second scenario assumes $\tau_d = 0$. For the first scenario, we observe three specific spectral domains: in the first one, $\mathrm{Im}(k_0) \gg \mathrm{Re}(k_0)$, where absorption dominates and for $\omega\tau_0 \to 0$, there is a convergence towards $k_0^s = \sqrt{\tau_d/\kappa_0}$ (purely imaginary wavenumber); in the second domain, one has $\mathrm{Re}(k_0) \gg \mathrm{Im}(k_0)$, i.e., propagative regime; and in the third domain one has $\mathrm{Re}(k_0) \approx \mathrm{Im}(k_0)$, i.e., the scattering and absorption are balanced. For the second scenario, we have the same behavior for spectral regions 2 and 3, and in region 1, we have $\mathrm{Re}(k_0) \approx \mathrm{Im}(k_0)$, and both converge towards 0 in the static regime, since there is no absorption there.

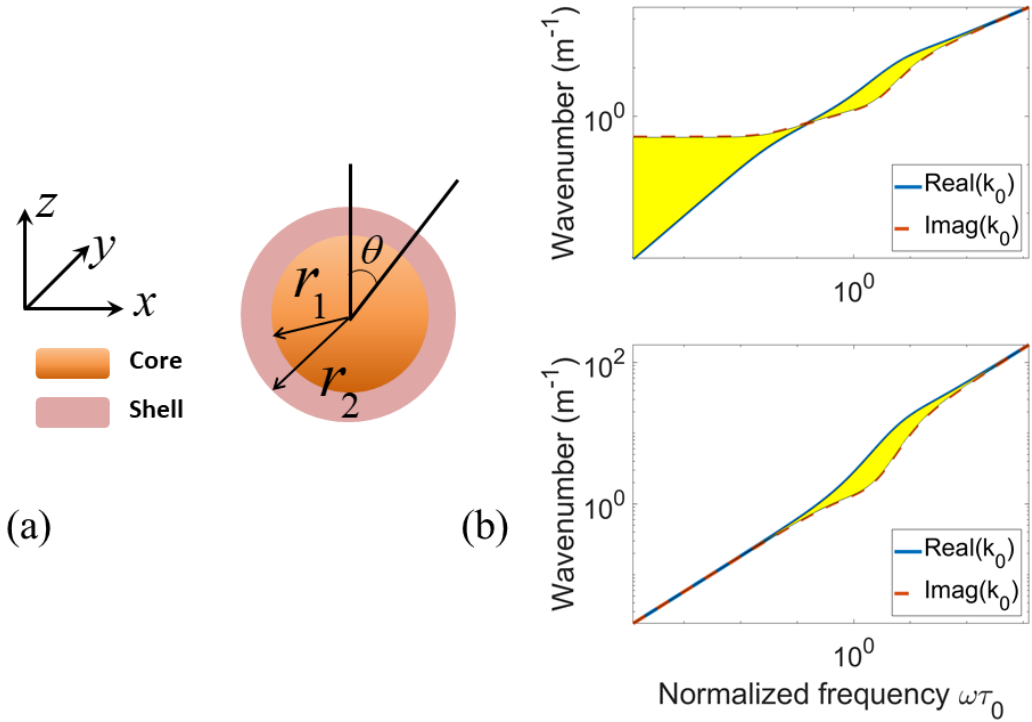

**Figure 1.** (a) Schematic representation of the core-shell system in cross-sectional view in the *x-z* plan. (b) Dispersion relation for the medium computed using Eq. (6) for the Maxwell

Cattaneo, with (top) and without (bottom) the extra absorption term (corresponding to diffusive waves).

## III. Scattering Cancellation for the Maxwell-Cattaneo Equation

### III.1. Set-up of the Scattering Problem

We now move to the derivation and analysis of the scattering problem in the framework of Maxwell-Cattaneo heat waves. We assume that incident waves are plane waves (which is a good approximation for spherical waves far away from the source). These plane waves impinge on the core-shell system (spherical object and shell, schematized in Fig. 1(a)) and shown in the *x-z* plane. The geometrical and physical parameters of the object and the cloak are identified using subscripts 1 and 2, respectively, while free-space parameters have subscript 0 ($k_0$ as wavenumber in free-space, for example). We further make the assumption that the origin of the spherical coordinates lies in the center of the object, without loss of generality. It is thus possible to find solutions to Eq. (5) by developing the temperature field in the different regions of space in terms of spherical Bessel and Hankel functions of different kinds. These eigensolution are shown to obey the Helmholtz equation with an equivalent (complex) wavenumber in the argument.

Inside free-space, two components of the temperature field (normalized to the incident amplitude, or equivalently assuming a unit-amplitude incident field) exist, i.e., the incident field $T^{\text{in}}$, i.e., the plane wave $e^{ik_0 r \cos\theta}$,

$$T^{\text{in}} = \sum_{l=0}^{\infty} i^l (2l+1) j_l(k_0 r) P_l(\cos\theta), \qquad (6)$$

with $j_l$ denoting the spherical Bessel function of order $l$, and $\theta$ the incidence angle of the waves, where we took advantage of the spherical symmetry of the problem, and placed the source in the $z$-axis. The second field in free-space, and the most important one, regarding our problem, is the scattered temperature field $T^{sc}$, resulting from the inhomogeneity encountered by the wave where it impinges from one medium to another, i.e.,

$$T^{sc} = \sum_{l=0}^{\infty} i^l (2l+1) s_l h_l^{(1)}(k_0 r) P_l(\cos\theta), \tag{7}$$

with $P_l$ the Legendre polynomial of order $l$ and $s_l$ are complex numbers accounting for the scattering coefficients, and $h_l^{(1)}$ are the spherical Hankel functions of order $l$ and the first kind.

Inside the object, i.e. for $r \leq a_1$, we have

$$T^1 = \sum_{l=0}^{\infty} i^l (2l+1) a_l j_l(k_1 r) P_l(\cos\theta), \tag{8}$$

and finally, inside the shell, i.e. for $r \in [a_1, a_2]$

$$T^2 = \sum_{l=0}^{\infty} i^l (2l+1) \{b_l j_l(k_2 r) + c_l y_l(k_2 r)\} P_l(\cos\theta), \tag{9}$$

where $a_l$, $b_l$, and $c_l$ are unknown coefficients that will be determined along with $s_l$ by applying the boundary conditions at $r = a_1$ and $r = a_2$. The boundary conditions involve continuity of the temperature field $T$, as well as its flux $\kappa_i \nabla T$, or more precisely its radial component $\kappa_r \partial_r T$.

By denoting the scattering coefficients in a more convenient form, i.e.,

$$s_l = \frac{-\varsigma_l}{\varsigma_l + i\beta_l}, \tag{10}$$

one finds that

$$\varsigma_l = \det \begin{pmatrix} -j_l(k_1 a_1) & y_l(k_2 a_1) & j_l(k_2 a_1) & 0 \\ 0 & y_l(k_2 a_2) & j_l(k_2 a_2) & y_l(k_0 a_2) \\ -\kappa_1 k_1 j_l'(k_1 a_1) & \kappa_2 k_2 y_l'(k_2 a_1) & \kappa_2 k_2 j_l'(k_2 a_1) & 0 \\ 0 & \kappa_2 k_2 y_l'(k_2 a_2) & \kappa_2 k_2 j_l'(k_2 a_2) & \kappa_0 k j_l'(k_0 a_2) \end{pmatrix}, \tag{11}$$

and

$$\beta_l = \det \begin{pmatrix} -j_l(k_1 a_1) & y_l(k_2 a_1) & j_l(k_2 a_1) & 0 \\ 0 & y_l(k_2 a_2) & j_l(k_2 a_2) & j_l(k_0 a_2) \\ -\kappa_1 k_1 j_l'(k_1 a_1) & \kappa_2 k_2 y_l'(k_2 a_1) & \kappa_2 k_2 j_l'(k_2 a_1) & 0 \\ 0 & \kappa_2 k_2 y_l'(k_2 a_2) & \kappa_2 k_2 j_l'(k_2 a_2) & \kappa_0 k y_l'(k_0 a_2) \end{pmatrix}. \tag{12}$$

In these determinants, the wavenumbers $k_j$, $(k = 0, 1, 2)$ are given by the dispersion relation

$$k_j = \omega \sqrt{\frac{\omega \tau_j + i}{\omega \kappa_j - i\omega^2 \tau_j \sigma_j}}, \tag{13}$$

where one ignores the extra absorption term corresponding to DPDWs.

**III.2. Ideal Cloaking Conditions**

Now, we are in a position to compute both analytically and numerically the scattering cross-section (SCS) of the system $\Sigma^{sc}$, which represents the quantity that can be measured by a radar device, placed far away from the structure. In a sense, this scalar physical parameter quantifies how *visible* (or equivalently *invisible*) an object or a collection of objects may be visible to external observer. In order to render an object completely opaque, one must consider instead the extinction cross-section that contains both scattering and absorption, but this goes beyond the scope of this study. The SCS is given by

$$\Sigma^{sc} = \frac{4\pi}{|k_0|^2} \sum_{l=0}^{\infty} (2l+1) \frac{|\varsigma_l|^2}{|\varsigma_l + i\beta_l|^2}. \tag{14}$$

To make an object invisible or transparent, it is straightforward to see that one needs to have $\Sigma^{sc} = 0$. But this is an impossible task since the SCS involves an infinite number of terms.

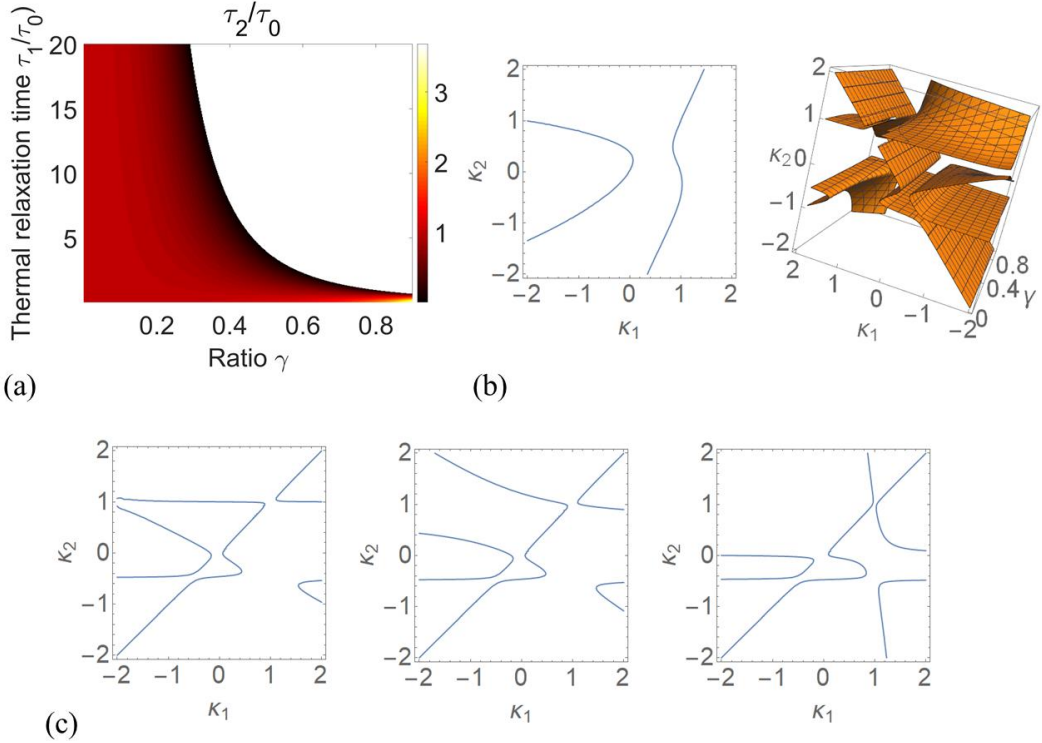

**Figure 2.** (a) Monopole scattering cancellation deduced from Eq. (15). (b) Dipole scattering cancellation for the imaginary and real parts deduced from Eq. (16). (c) Cross-sectional plots for $\gamma = 0.01, \ 0.5, \ \text{and} \ 0.95$, respectively (from left to right).

A more realistic task is to cancel the first few leading order terms present in Eq. (7). This does not render an object invisible, but in the quasi-static limit, this substantially reduces scattering and is a viable way to realize near-perfect invisibility, since only the first (dominant) terms contribute non-negligibly to the scattering. When the dimension of the system is smaller than the wavelength, $ka_2 \ll 1$, cancelling the first two terms is sufficient, as $\Sigma^{sc} \approx 4\pi/|k_0|^2 \left(|s_0|^2 + 3|s_1|^2\right)$.

$$\frac{\tau_2 - \tau_0}{\tau_2 - \tau_1} - \gamma^3 = 0, \tag{15}$$

with the ratio $\gamma = a_1/a_2$, for $\varsigma_0 = 0$, and

$$\frac{\left[(\kappa_0 - \kappa_2) - i\omega(\tau_0\sigma_0 - \tau_2\sigma_2)\right]\left[(\kappa_1 + 2\kappa_2) - i\omega(\tau_1\sigma_1 + 2\tau_2\sigma_2)\right]}{\left[(\kappa_1 - \kappa_2) - i\omega(\tau_1\sigma_1 - \tau_2\sigma_2)\right]\left[(\kappa_0 + 2\kappa_2) - i\omega(\tau_0\sigma_0 + 2\tau_2\sigma_2)\right]} - \gamma^3 = 0, \tag{16}$$

for $\varsigma_1 = 0$.

Figure 2(a) plots the solution of Eq. (15), for varying ratio $\gamma$ and thermal relaxation time $\tau_1$ (in units of $\tau_0$). One can observe a classical behavior in its lower part, for $k_0 a_1 = 0.5$. The white region corresponds to negative values of $\tau_2$, and is henceforth ignored since it is unphysical, whereas the red part corresponds to values of $\tau_2$ in the range of $0.1\tau_0$ and $3.5\tau_0$. The upper black curve stands for $\tau_2 = 0$ (i.e. $\gamma^3 \tau_1/\tau_0 = 1$). Equation (16) is however different from other scenarios (heat waves of DPDWs) as its first term is complex-valued. To realize perfect dipole scattering cancellation, one has to cancel both its real and imaginary parts. However, by decomposing it, one can observe that the imaginary part takes close-to-zero values, except for $\kappa_2 = \kappa_1$ and $\kappa_2 = -1/2\kappa_0$. So it is possible to cancel the real part of Eq. (16) and exclude these values ($\kappa_1$ and $-1/2\kappa_0$) from the possible solutions. The real part of the left-hand-side of Eq. (16) can be cast in the form of

$$A/B - \gamma^3 = 0 \tag{17}$$

with,

$$A = (x_1 - x_2)(x_3 + x_4)(x_5 - x_6)(x_7 + x_8), \tag{18}$$

and

$$B = (x_3^2 + x_4^2)(x_7^2 + x_8^2), \tag{19}$$

where $x_1 = \kappa_0 - \kappa_2$, $x_2 = \omega(\tau_0\sigma_0 - \tau_2\sigma_2)$, $x_3 = \kappa_1 - \kappa_2$, $x_4 = \omega(\tau_1\sigma_1 - \tau_2\sigma_2)$, $x_5 = \kappa_1 + 2\kappa_2$, $x_6 = \omega(\tau_1\sigma_1 + 2\tau_2\sigma_2)$, $x_7 = \kappa_0 + 2\kappa_2$, and $x_8 = \omega(\tau_0\sigma_0 + 2\tau_2\sigma_2)$.

Equation (17) is of fourth-order, so one expects to have four possible solutions for $\kappa_2$. This is exactly what one can observe from Fig. 2(b), 2(c), where four branches can be distinguished from the three-dimensional plot. Figure 2(c) plots the solutions (contour plots) for different values of the ratio $\gamma$, i.e. 0.01, 0.5, and 0.95, from left to right, respectively. For each value of $\kappa_1$ (and $\gamma$), one has four possible solutions for $\kappa_2$ that cancel Eq. (17).

### III.3. Scattering Cancellation Technique

The considered geometry is shown in Fig. 1(a), with radius $a_1 = 1$ cm and we choose the radius of the shell $a_1 = 1.15$ cm (or $\gamma = 0.87$), which corresponds to a relatively thin cloaking shell. The parameters of the object are $\kappa_1 = 3\kappa_0$ and $\tau_1\sigma_1 = \tau_0\sigma_0$. Then, the SCS is numerically computed for a finite scattering order $N_0$. In fact, it was shown in many recent studies that the scattering coefficients $|s_l|$ are of the order $|k_0a_1|^{2l+1}$, so when $|k_0a_1| \ll 1$, only a few coefficients is enough to reach convergence of $\Sigma^{sc}$. In this study, one verifies convergence, and choose $N_0 = 10$. The SCS of the structure is thus normalized with the SCS of the bare object and plotted in logarithmic scale in Fig. 3(a) versus the diffusivity coefficient $\kappa_2$ (in units of $\kappa_0$) and the thermal relaxation time $\tau_2$ (in units of $\tau_0$). It can be seen from this two-dimensional (2D) plot, that the normalized SCS is increased in some regions (dark red color) which corresponds to enhanced scattering, as well as reduced scattering (dark blue color) which corresponds to transparency (or cloaking) effect. In particular, a minimum of $\Sigma_2^{sc}$ is obtained around the values $\kappa_2 = 0.6\kappa_0$ and $\tau_2 = 0.3\tau_0$ (white dot in the Fig. 3(a)). To isolate the effect of $\kappa_2$ and $\tau_2$ on the scattering reduction, a plot of $\Sigma_2^{sc}$ is given versus $\kappa_2$ for different values of $\tau_2$. One can see that two regimes exist. First, for small values of $\tau_2$, no reduction is significantly obtained for all values of $\kappa_2$ (red line on the bottom of Fig. 3(a)). Then, the scattering reduction regime starts operating, and is maximal for $\tau_2 = 0.3\tau_0$, and again starts disappearing gradually. This shows that effective cloaking can be obtained, only by optimizing both parameters ($\tau_2$ and $\kappa_2$) of the shell, unlike in static studies, where only $\kappa_2$ is considered. One goes one step further by analyzing the SCS in polar coordinates, by imposing the values of $\kappa_2$ and $\tau_2$ that better reduce the SCS and plotting $\Sigma_2^{sc}$ in the far-field versus the angle of observation $\theta$. Figure 3(c) shows such plot, and significant reduction of $\Sigma_2^{sc}$ is obtained for all angles, by using the cloaking shell. The near-field plot of the temperature field, is further given in Fig. 3(d), in the presence of the same core-shell structure for $k_0 a_1 = 0.5$, and shows no perturbation of the amplitude of the field in the vicinity of the cloaked object, thus demonstrating the robustness of the SCS for such kind of waves.

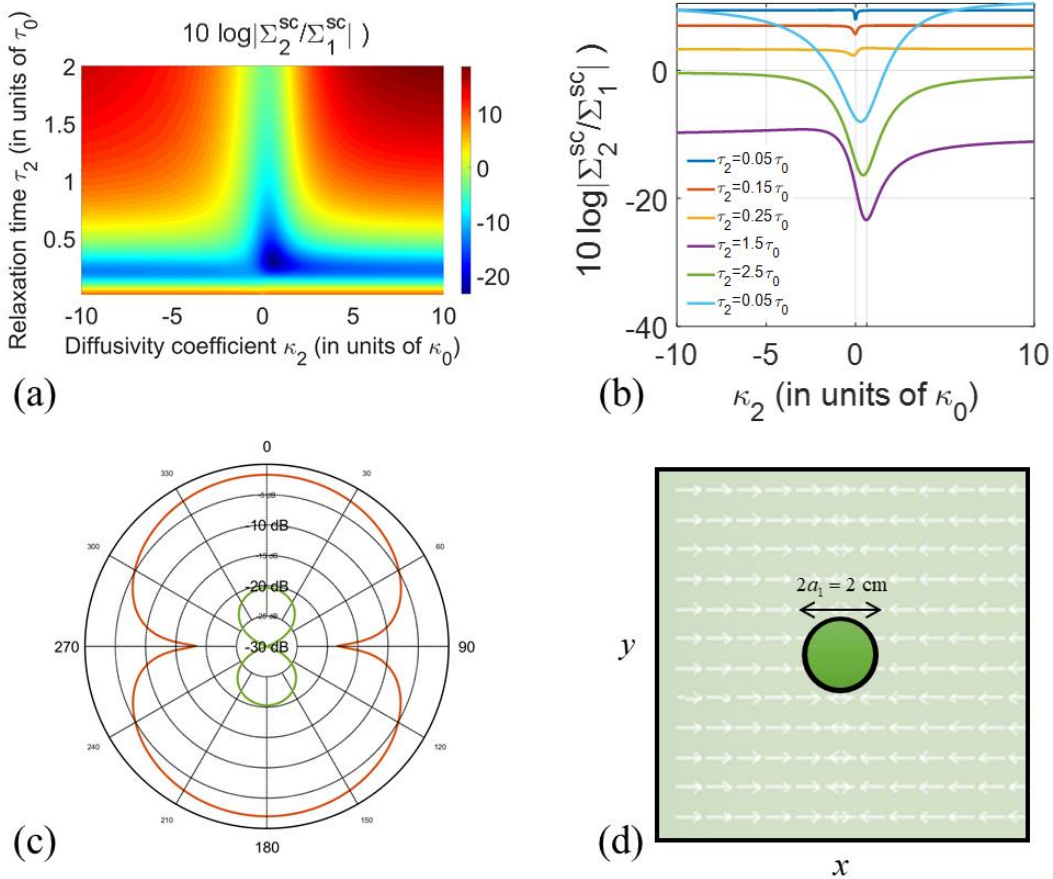

**Figure 3.** (a) SCS in logarithmic scale versus the normalized diffusivity coefficient $\kappa_2$ and thermal relaxation time $\tau_2$. (b) Cross-sectional plots from the two-dimensional graph (in Fig. 3(a)) for different values of $\tau_2$. (c) Far-field polar scattering amplitude for both the uncloaked object (red line) and cloaked one (green line). Near-field plot of the temperature field in the vicinity of the core-shell structure.

## IV. Mantle Cloaking

Mantle cloaking was shown to make possible considerable scattering reduction for several kinds of waves (7, 24, 25, 30) using an ultrathin metasurface cloak. It relies on coating an object with a metasurface (ideally 2D) with tailored impedance to cancel the radar scattering. In the case of Maxwell-Cattaneo heat waves, the boundary conditions ate the interface $r = a_1$ is the same as in Section C (i.e. continuity of $T$ and its flux $\kappa_r \partial_r T$). However, at the boundary $r = a_2$ (i.e. at the metasurface), one has $T(\mathrm{r} = a_2^-) = T(\mathrm{r} = a_2^+)$, where the -,+ signs stand for the inside, outside the metasurface, respectively, as schematized in Fig. 4(a). The second condition is

$$(\kappa_0 - i\omega\sigma_0\tau_0)\frac{\partial T}{\partial r}(r = a_2^+) - (\kappa_2 - i\omega\sigma_2\tau_2)\frac{\partial T}{\partial r}(r = a_2^-) = Z_{\mathrm{MC}}^{-1} T(a_2). \qquad (20)$$

With $Z_{\mathrm{MC}}$ the average surface impedance of heat waves relating the temperature $T$ to its flux $\kappa_r \partial_r T$. The scattering coefficients in this configuration become

$$\varsigma_l = \det \begin{pmatrix} -j_l(k_1 a_1) & y_l(k_0 a_1) & j_l(k_0 a_1) & 0 \\ 0 & y_l(k_0 a_2) & j_l(k_0 a_2) & y_l(k_0 a_2) \\ -\kappa_1 k_1 j_l'(k_1 a_1) & \kappa_0 k_2 y_l'(k_0 a_1) & \kappa_0 k_2 j_l'(k_0 a_1) & 0 \\ 0 & y_l'(k_0 a_2) + \chi y_l(k_0 a_2) & j_l'(k_0 a_2) + \chi j_l(k_0 a_2) & j_l'(k_0 a_2) \end{pmatrix},$$

$$(21)$$

with $\chi = i\omega/(Z_{MC}k_0(\kappa_0 - i\omega\sigma_0\tau_0))$ a dimensionless function. For $l=0$, and by imposing $\varsigma_0 = 0$, we get

$$X_{MC} = \frac{2}{3\gamma^3 \omega a_1}\left(\kappa_0\gamma^3 + \text{Re}\left[\frac{(\kappa_0 - i\omega\sigma_0\tau_0) + 2(\kappa_1 - i\omega\sigma_1\tau_1)}{(\kappa_1 - i\omega\sigma_1\tau_1) - (\kappa_0 - i\omega\sigma_0\tau_0)}\right]\right), \qquad (22)$$

by denoting $X_{MC} = \text{Im}[Z_{MC}]$ the reactive part of the total impedance.

Figure 4(b) gives the SCS of the structure shown in Fig. 4(a), i.e. the object surrounded by the mantle cloak with $a_1 = 1$ cm and $a_2 = 1.15$ cm. It can be clearly seen that a substantial scattering reduction can be achieved around the design wavenumber $k_0 a_1 = 0.5$ (that corresponds to $\omega_0$).

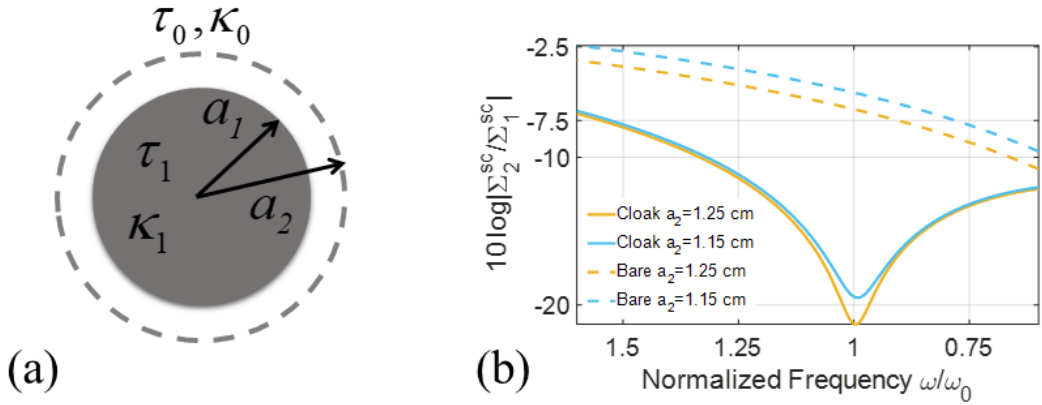

**Figure 4.** (a) Scheme of the object surrounded by the metasurface cloak of radius $a_2$. (b) SCS in logarithmic scale versus the normalized frequency $\omega/\omega_0$ for the object and cloaked object for two different radii of the metasurface.

## V. Conclusions

In summary, this study presented the first demonstration of scattering cancellation cloaking for heat waves obeying the Maxwell-Cattaneo transfer law, which permits to avoid unphysical features associated with Fourier transfer law, such as instantaneous diffusion. Both SCT and mantle cloaking were analyzed for this kind of waves and shown to result in near-perfect invisibility, in all directions. The far-field and near-field numerical calculations demonstrate the robustness of such thin cloaks, which are also easy to fabricate, due to the isotropy and homogeneity of the cloaking shells, unlike TO-based cloaks.

## Acknowledgement

S.G. wishes to thank a visiting position in the group of Prof. R.V. Craster at Imperial College London funded by EPSRC program grant "Mathematical fundamentals of Metamaterials for multiscale Physics and Mechanics" (EP/L024926/1).